\documentclass[11pt,a4paper]{emulateapj}
\usepackage{graphics}
\usepackage{epsfig}
\usepackage{natbib}
\usepackage{amsmath}
\usepackage{textcomp}
\usepackage{threeparttable}
\usepackage{appendix}
\usepackage{longtable}

\begin{document}

\title{Modeling the HD32297 Debris Disk with Far-IR Herschel Data}

\author{J. K. Donaldson\altaffilmark{1},  J. Lebreton\altaffilmark{2}, A. Roberge\altaffilmark{3}, J.-C. Augereau\altaffilmark{2},  A. V. Krivov\altaffilmark{4}}

\altaffiltext{1}{Department of Astronomy, University of Maryland, College Park, MD 20742; \texttt{jessd@astro.umd.edu}}
\altaffiltext{2}{UJF - Grenoble 1 / CNRS-INSU, Institut de Plan\'{e}tologie et d'Astrophysique de Grenoble (IPAG) UMR 5274, Grenoble, F-38041, France}
\altaffiltext{3}{Exoplanets and Stellar Astrophysics Laboratory, NASA Goddard Space Flight Center, Code 667, Greenbelt, MD 20771}
\altaffiltext{4}{Astrophysikalishes Institut, Friedrich-Schiller-Universit\"{a}t Jena, Schillerg\"{a}\ss{}chen 2-3, 07745 Jena, Germany}

\begin{abstract}
HD32297 is a young A-star ($\sim30$ Myr) 112 pc away with a bright edge-on debris disk that has been resolved in scattered light.  We observed the HD32297 debris disk in the far-infrared and sub-millimeter with the {\it Herschel Space Observatory} PACS and SPIRE instruments, populating the spectral energy distribution (SED) from 63 to 500$\,\mu$m.  We aimed to determine the composition of dust grains in the HD32297 disk through SED modeling, using geometrical constraints from the resolved imaging to break degeneracies inherent in SED modeling.  We found the best fitting SED model has 2 components: an outer ring centered around 110 AU, seen in the scattered light images, and an inner disk near the habitable zone of the star.  The outer disk appears to be composed of grains $>2\,\mu$m consisting of silicates, carbonaceous material, and water ice with an abundance ratio of 1:2:3 respectively and 90\% porosity. These grains appear consistent with cometary grains, implying the underlying planetesimal population is dominated by comet-like bodies.  We also discuss the $3.7\sigma$ detection of [\ion{C}{2}] emission at $158\,\mu$m with the {\it Herschel} PACS Spectrometer, making HD32297 one of only a handful of debris disks with circumstellar gas detected.  
\end{abstract}

\section{Introduction}
Debris disks are circumstellar disks composed of dust produced during collisions of planetesimals.  In the youngest disks (10-100 Myrs), the planetesimals may deliver volatiles such as water to still-forming terrestrial planets.  Although the planetesimals themselves are undetectable, the dust they produce radiates thermally in the infrared (IR) and sub-millimeter (mm) and scatters starlight at optical and near-IR wavelengths.  These grains  provide clues to the composition of their parent bodies.  

The outer regions of several debris disks have been imaged in scattered light.  Resolved images can constrain the morphology of the disk's outer regions, but the composition of their grains cannot be uniquely determined from these images alone.  Mid-IR spectra - useful for determining grain composition in younger protoplanetary disks - also fail to provide constraints on the dust composition in most debris disks; by this point, the remaining grains are too large to emit solid state features in the mid-IR \citep{Chen06}.  While there are a few notable exceptions, most debris disks require modeling of the full spectrum, from optical to mm wavelengths, to probe the grain composition.

The {\it Herschel Space Observatory} \citep{Pilbratt10} was launched in May 2009, presenting a new opportunity for sensitive far-IR and sub-mm observations.  The PACS \citep{Poglitsch10} and SPIRE \citep{Griffin10} instruments have photometric and spectroscopic capabilities spanning a wavelength range of $\sim 60-500\,\mu$m.  These data are crucial for detailed modeling of a disk's spectral energy distribution (SED) because they span the wavelength range where the thermal emission from debris disks typically peaks.  

Unfortunately, SED modeling of debris disks is hampered by degeneracies in the models between disk geometry and grain properties.  Thankfully, resolved imagery can be used to break some of the degeneracies by providing geometrical constraints.  We have used the {\it Herschel Space Observatory}'s PACS and SPIRE instruments to obtain far-IR and sub-mm photometry and spectroscopy of the disk surrounding the $\sim 30$ Myr-old \citep{Kalas_30Myr} A-star, HD32297.  This edge-on disk has been imaged several times in scattered light, thereby constraining the disk geometry. The {\it Herschel} observations fill in a large gap in the SED, which allows us to model the grain composition in more detail.  

The HD32297 disk \citep[112 pc away;][]{vanleeuwen07} was first resolved in the near-IR out to a distance of $3.3^{\prime\prime}$ (400 AU) from the star with HST NICMOS \citep{Schneider05} and later resolved at several other near-IR wavelengths \citep{Debes09,Mawet09}.  Recently, Angular Differential Imaging (ADI) has been used to resolve the disk in the near-IR with ground-based facilities \citep{Currie12, Boccaletti12}.  Additionally, the disk has been marginally resolved at mid-IR \citep{Moerchen07, Fitzgerald07} and millimeter wavelengths \citep{Maness08}.  

The HD32297 debris disk has a few unique features; one of the more luminous debris disks (L$_\text{IR}/L_\ast \sim 10^{-3}$), HD32297 is also one of only a handful of debris disks where circumstellar gas has been detected.  \cite{Redfield07} detected \ion{Na}{1} in absorption towards HD32297 that was not found towards any neighboring stars.  The additional peculiarity of brightness asymmetries and warping seen in scattered light images were analyzed by \cite{Debes09}, who concluded that these features could be caused by the disk's motion through the interstellar medium (ISM).  

The {\it Herschel} data presented here were acquired as part of the  {\it Herschel} Open Time Key Programme entitled ``Gas in Protoplanetary Systems'' \citep[GASPS; ][]{Dent12}.  The PACS data were taken as part of the main program, and the SPIRE data were taken as part of an Open Time proposal to follow up GASPS debris disks at longer wavelengths (OT2\_aroberge\_3; PI: A.\ Roberge).  Here we present the PACS and SPIRE observations of HD32297 as well as the results of the modeling of the entire SED.  In Section 2, we present the data and describe the data reduction.  In Section 3, we describe the SED modeling and we discuss the results in Sections 4 and 5.  

\vspace{0.3cm}

\section{Observations and Data Reduction}

Observations of HD32297 at 70, 100, and 160$\,\mu$m were taken with the {\it Herschel} PACS instrument in ScanMap mode with the medium scan speed of 20$^{\prime\prime}$s$^{-1}$.  The 70$\,\mu$m observations were taken at a scan angle of $63\degr$ and consisted of 8 scan legs with scan lengths of 3$^{\prime}$, and 2$^{\prime\prime}$ separation between the legs, for a total observing time of 220 seconds.  The 100 and 160$\,\mu$m observations were taken simultaneously with slightly larger maps of 10 legs with the same leg length and separation as the 70$\,\mu$m observation.  Data at 100 and 160$\,\mu$m were taken at two different scan angles, 70 and $110\degr$, with a total duration of 276 seconds per scan angle.  The observations at the two scan angles were then combined to reduce noise due to streaking in the scan direction.  
The data were reduced with HIPE 8.2 \citep{Ott10} using the standard reduction pipeline.  The final maps were chosen to have a pixel scale corresponding to the native pixel scale of the PACS detectors, 3.2$^{\prime\prime}$ for the 70 and 100$\,\mu$m images and 6.4$^{\prime\prime}$ for the 160$\,\mu$m images.  

Simultaneous images at 250, 350, and 500$\,\mu$m of HD32297 were taken with the SPIRE instrument.  The observations were made in the Small Scan Map mode with a scan speed of 30$^{\prime\prime}$s$^{-1}$, with two repetitions and a total observation time of 307 seconds.  The data were reduced with HIPE 8.2, producing final maps with pixels scales of 6, 10, and 14$^{\prime\prime}$ for the 250, 350, and 500$\,\mu$m images respectively.  HIPE produces SPIRE images with units of Jy beam$^{-1}$, which we converted into Jy pixel$^{-1}$ for analysis using beam areas of 423, 751, and 1587 arcsec$^2$ for the 250, 350, and 500$\,\mu$m images respectively.  

The {\it Herschel} PACS spectroscopy of HD32297 was taken in two modes, the lineSpec and rangeSpec modes.  The lineSpec observations targeted the [\ion{O}{1}] 63.2$\,\mu$m line with a duration of 3316 seconds.  The rangeSpec observations targeted six lines, [\ion{O}{1}] at 145.5$\,\mu$m, [\ion{C}{2}] at 157.7$\,\mu$m, two o-H$_2$O lines at 78.7 and 179.5$\,\mu$m, and 2 CO lines at 72.8 and 90.2$\,\mu$m.  The total rangeSpec observation time was 5141 seconds, divided into three observing segments, covering the six lines two at a time.  A deep follow-up rangeSpec observation was performed targeting just the [\ion{C}{2}] 157.7$\,\mu$m line with a duration of 4380 seconds.  
The spectroscopic data were reduced using HIPE 8.2.  An aperture correction is applied in HIPE to account for point source flux loss. The spectra were produced with a pixel scale corresponding to the native resolution of the instrument.  

\vspace{0.5cm}
\section{Analysis}
\subsection{Herschel PACS photometry}
The HD32297 disk is a spatially unresolved point source to {\it Herschel}. The full-width at half-maximum of the {\it Herschel} PACS beam at $70\,\mu$m is $5.6^{\prime\prime}$, so the bulk of the thermal emission must be within $\sim 300$ AU. This is consistent with resolved images which suggest the disk peaks at $\sim110$ AU \citep{Debes09,Currie12, Boccaletti12}.   Aperture photometry was performed with apertures of 12$^{\prime\prime}$ for the 70 and 100$\,\mu$m images and 22$^{\prime\prime}$ for the 160$\,\mu$m image.  

Aperture corrections provided by the {\it Herschel} PACS ICC\footnotemark[1] were applied, but color corrections were not applied.  The uncertainty in the flux was calculated from the standard deviation of the sky background in an annulus around the aperture.  The annulus was placed between 20-30$^{\prime\prime}$ from the central star for the 70 and 100$\,\mu$m images and between 30-40$^{\prime\prime}$ for the 160$\,\mu$m image.  An absolute calibration error of 2.64, 2.75, and 4.15\%\footnotemark[1] for the 70, 100, and 160$\,\mu$m images respectively was added in quadrature to the uncertainty measured from the sky background.  Results from the {\it Herschel} PACS photometry observations are given in Table \ref{tab:photometry}.   

\footnotetext[1]{PICC-ME-TN-037: \url{http://herschel.esac.esa.int/twiki/pub/Public/\newline PacsCalibrationWeb/pacs\_bolo\_fluxcal\_report\_v1.pdf}}

\begin{table}[ht]
\centering
\caption{{\it Herschel} PACS and SPIRE photometry results \label{tab:photometry}}
\begin{tabular}{l c c}
\tableline \tableline
Obs.\ Id	& Wavelength	& Flux $\pm$ Error	\\
		& ($\mu$m)	& (Jy)	\\
\tableline 
1342193125	& 70			& $1.038\pm0.029$	\\
1342217452-3	& 100		& $0.770\pm0.022$	\\
1342217452-3	& 160		& $0.403\pm0.020$	\\
1342240033	& 250		& $0.153\pm0.012$	\\
1342240033	& 350		& $0.071\pm0.008$		\\
1342240033	& 500		& $0.045\pm0.007$		\\
\tableline 
\end{tabular}
\end{table}

\subsection{Herschel SPIRE photometry} 

Aperture photometry was performed with aperture radii of 22, 30, and 42$^{\prime\prime}$ for the 250, 350, and 500$\,\mu$m images respectively.  The sky background was estimated from an annulus with radius of 60-90$^{\prime\prime}$ from the central star and subtracted from the measured flux.  Aperture corrections were applied according to the SPIRE data reduction guide\footnotemark[2]. A color correction\footnotemark[2] of $\sim5$\% was applied assuming a Rayleigh-Jeans law slope of $F_\nu \propto \nu^2$.  This correction could be off by 5\% if the slope of the SED in the sub-mm is steeper than Rayleigh-Jeans.  The uncertainties in the flux measurements come from the standard deviation of the sky background added in quadrature with a 7\% absolute calibration error.  Results from the {\it Herschel} SPIRE photometry are listed in Table \ref{tab:photometry}.

\footnotetext[2]{SDRG 5.7: \url{http://herschel.esac.esa.int/hcss-doc-9.0/}}

\subsection{Herschel PACS spectroscopy}

The PACS spectrometer is an Integral Field Unit (IFU) spectrometer that has a $5\times5$ array of spaxels with each spaxel covering a $9.4^{\prime\prime}\times9.4^{\prime\prime}$ region.  We verified that the star was well-centered on the central spaxel during the observations by comparing the observations to a model of the transmission of a theoretical PSF through the PACS IFU.  We shifted the model PSF and calculated the fractional flux in the different spaxels as a function of offset.  Comparison to observations of HD32297 indicate the star is not significantly offset.  This method is the same as the one used by Howard et al.\ (in preparation) for PACS observations of Taurus that had pointing errors.  

We use data from the central spaxel only for analysis.  No lines were detected in the lineSpec or the first rangeSpec observation.  We calculated continuum flux values by fitting straight lines to the data, and the results are listed in Table \ref{tab:spec}. We excluded five pixels on each edge from the continuum fit because PACS spectra have enhanced noise at the edges.  Emission in the [\ion{C}{2}] 157.7$\,\mu$m line was seen in the second deep rangeSpec observation.  The continuum flux was found by fitting a line to the spectrum while masking out 0.5$\,\mu$m around the line center.  The rms noise was also calculated in the region surrounding the line.  The number of pixels used to calculate the noise was chosen such that the signal-to-noise in the emission line was maximized, i.e.\ the rms noise was minimized.  

\begin{figure}
\hspace{-2.5em}
\epsscale{1.25}
\plotone{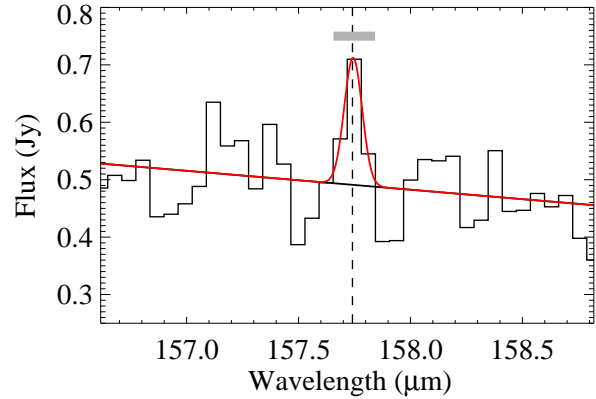}
\caption{The [\ion{C}{2}] 157.7$\,\mu$m emission line from the HD32297 debris disk.  The dashed line represents the expected position of the [\ion{C}{2}] line.  The continuum was fit with a straight line (black solid line), and the emission line was fit with a Gaussian (red solid line).   The integrated line flux was measured from the 3 pixels surrounding the line (indicated by the grey bar). The integrated line flux is 3.7$\sigma$ above the continuum.   \label{fig:CIIline}}
\end{figure}

The line flux was integrated over the three pixels surrounding the line (marked by the grey bar in Figure \ref{fig:CIIline}).  The uncertainty in the line flux was found by propagating the rms noise.  Upper limits on the other lines were found using the same method of error propagation at the expected line center.  These values are reported in Table \ref{tab:spec}.  The integrated line flux of the [\ion{C}{2}] line is detected at the 3.7$\sigma$ level above the continuum in the central spaxel from the deep rangeSpec observation.  All 25 spaxels were searched for a significant [\ion{C}{2}] line signal, and no other emission was found with more than a $2\sigma$ significance.  Since the line is present only in the central spaxel, we believe the gas is associated with the star rather than from the surrounding ISM.  

\begin{table}[ht]
\begin{center}
\caption{{\it Herschel} PACS Spectroscopy Results \label{tab:spec}}
\begin{tabular}{l r r c}
\tableline
Line & \hspace{-1cm}Wavelength 	& Continuum 	& Integrated Line Flux\tablenotemark{a}	\\
Name	  & ($\mu$m) 	& Flux (Jy)  		&  $\times 10^{-18}$ (W/m$^2$)	\\
\tableline
\ion{O}{1} 63	  & 63.184	& $1.28\pm0.20$		& $<7.29$	\\
CO 72	  & 72.843	& $0.83\pm0.23$		& $<7.77$	\\
H$_2$O 79	  & 78.741	& $1.03\pm0.23$		& $<9.12$	\\
CO 90	  & 90.163	& $1.05\pm0.29$		& $<8.13$	\\
\ion{O}{1} 145	  & 145.525	& $0.55\pm0.09$		& $<3.99$	\\
\ion{C}{2} 158	(1) \tablenotemark{b} & 157.741	& $0.53\pm0.13$		& $<4.32$	\\
H$_2$O 180	  & 179.741	& $0.44\pm0.15$		& $<3.57$	\\
\ion{C}{2} 158 (2) \tablenotemark{b}& 157.741	& $0.50\pm0.06$		& $2.68\pm0.72$ \\
\tableline
\end{tabular}
\tablenotetext{1}{Non-detections are reported as $3\sigma$ upper limits}
\tablenotetext{2}{\ion{C}{2} 158 line was observed twice.  (1) - first rangeSpec observation (2) - second deeper rangeSpec observation}
\end{center}
\end{table}

\subsection{Column density of \ion{C}{2} in HD32297}

With the flux of the [\ion{C}{2}] line, we can determine the column density of \ion{C}{2} for comparison to the \ion{Na}{1} column density found by \cite{Redfield07}.  The [\ion{C}{2}] 158$\,\mu$m line arises from the transition between the two fine structure lines of the electronic ground state.  Since the next electronic energy level is three orders of magnitude higher, we can assume a two level population with a high accuracy.  Following \cite{Roberge11}, the column density is \begin{equation} N_\text{CII} = \frac{4\pi\lambda}{hc} \frac{F_{10}}{A_{10} x_{1} \Omega}, \end{equation} where the indices 1 and 0 indicate the upper and lower fine structure levels respectively, $\lambda = 157.7\,\mu$m, $F_{10} = 2.68\times10^{-18}$ W m$^{-2}$ is the integrated line flux, $A_{10}=2.4\times10^{-6}$ s$^{-1}$ is the spontaneous transition probability, $\Omega=0.357$ arcsec$^2$ is the angular source size estimated from resolved scattered light images, and $x_1$ is the fractional population of the upper level.  Assuming local thermodynamic equilibrium (LTE), $x_1$ is dependent on the excitation temperature.  \cite{Roberge11} give this as \begin{equation}x_1 = \frac{(2J_1+1)g_1 e^{-E_1/kT_{ex}}}{Q(T_{ex})},\end{equation} where $J_1=3/2$ is the angular momentum quantum number of the upper level, $g_N$ are the nuclear statistical weights for the two levels ($g_0 = 2$ \& $g_1=4$), $E_1/k = 91.21$ K is the energy difference between the two levels, $T_{ex}$ is the excitation temperature, and $Q(T_{ex})$ is the partition function, which here can be approximated in a two level system by \begin{equation} Q(T_{ex}) = g_0 + g_1 e^{-E1/kT_{ex}} .\end{equation}  

\begin{figure}
\hspace{-2.5em}
\epsscale{1.25}
\plotone{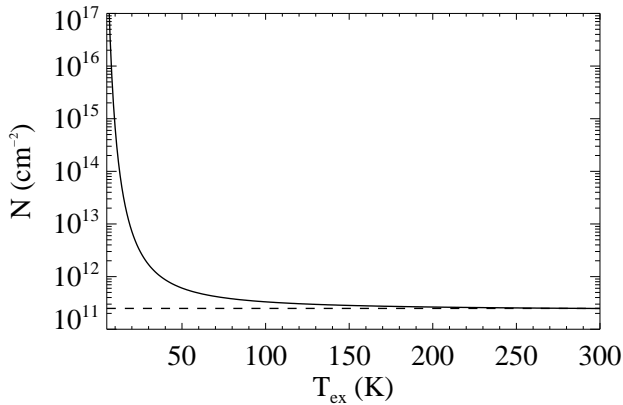}
\caption{Column density of \ion{C}{2} as a function of excitation temperature (solid line).  The dashed line shows the lower limit on the column density (N = $2.5\times10^{-11}$ cm$^{-2}$) for excitation temperatures $<300$ K.  \label{fig:NofTex}}
\end{figure}

With only one gas line, we cannot measure the excitation temperature.  However, we can derive a lower limit on the column density.  Figure \ref{fig:NofTex} shows the dependence of the column density on excitation temperature. From 1 to 300 K, we find a lower limit on the \ion{C}{2} column density of N$_{\text{CII}} > 2.5\times10^{11}$ cm$^{-2}$.  This value is similar to the column density of \ion{Na}{1} calculated by \cite{Redfield07} (N$_{\text{NaI} }= 2.5\times10^{11}$ cm$^{-2}$).  

\section{SED Modeling}
\subsection{SED data}

In addition to the {\it Herschel} data, we collected archive data for use in our SED modeling (see Tab. \ref{tab:archive}).  To constrain the stellar photosphere, we used {\it Hipparcos} B \& V \citep{Perryman97}, {\it 2MASS} J, H, \& K$_\text{s}$ \citep{Cutri03}, and {\it WISE} bands 1 \& 2 \citep{Wright10}. For the infrared excess, we used {\it IRAS} 25 \& 60$\,\mu$m \citep{Moor06}, {\it WISE} bands 3 \& 4 \citep{Wright10}, {\it Spitzer} MIPS 24$\,\mu$m \citep{Maness08}, and {\it Spitzer} IRS data taken from the enhanced data product in the {\it Spitzer} Heritage Archive.\footnotemark[3] We also included millimeter data from \cite{Meeus12}: the 1.2 mm flux from the MAMBO2 bolometer array on the IRAM 30 m telescope and the 1.3 mm flux from the Sub-Millimeter Array (SMA).  The uncertainties for these last two measurements (reported in Table \ref{tab:archive}) include 15\% calibration uncertainties added in quadrature.  

\footnotetext[3]{\url{irsa.ipac.caltech.edu/applications/Spitzer/SHA/}}

\begin{figure}
\hspace{-3em}
\epsscale{1.25}
\plotone{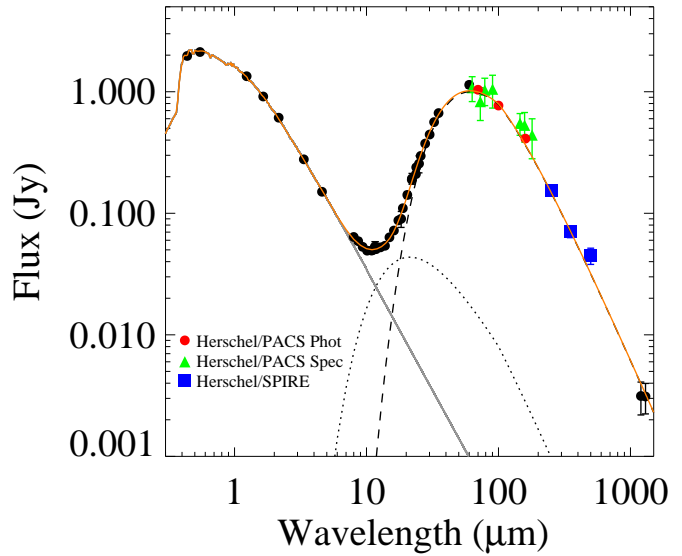}
\caption{Spectral energy distribution of HD32297 with simple blackbody fits.  The grey line is the best-fitting stellar photosphere model with T = 7750 K (see Section 4.2).  The data points that were taken from the literature and listed in Table \ref{tab:archive} are plotted here as well as the new PACS and SPIRE data given in Tables \ref{tab:photometry} and \ref{tab:spec}.   A two temperature modified blackbody model is over plotted (orange solid line), with the individual components also shown - a warmer blackbody with T = 240 K (dotted line) and a colder blackbody with T = 83 K (dashed line) as described in Section 4.4.  
  \label{fig:SED}}
\end{figure}

The HD32297 disk is also marginally resolved at mid-IR and millimeter wavelengths.  We used the total flux from Gemini North Michelle imaging at 11.2$\,\mu$m \citep{Fitzgerald07}, and Gemini South T-ReCS imaging at 11.7 \& 18.3$\,\mu$m \citep{Moerchen07}. We used the SMA flux at 1.3 mm rather than the total flux from the CARMA 1.3 mm resolved image \citep{Maness08}, because the unresolved data from the SMA has a smaller uncertainty.  
The photometric data used are listed in Table \ref{tab:archive} and plotted in Figure \ref{fig:SED}.  

We combined these data with our {\it Herschel} PACS and SPIRE photometry and the continuum values from the PACS spectroscopy.  Our data points fill in a large gap in the SED from 60 to 500$\,\mu$m (see Figure \ref{fig:SED}) and allow us to assess where the peak of the thermal emission is located. 

\begin{deluxetable}{r c l r}
\tablecolumns{4}
\tabletypesize{\scriptsize}
\tablecaption{Additional data used in SED modeling \label{tab:archive}}
\tablehead{Wavelength	& Flux \& Uncertainty	& Instrument & \hspace{-1cm}Reference	\\
($\mu$m)	& (Jy) & }
\startdata
     0.438 & $     1.952  \pm  0.026  $   &  {\it Hipparcos} B &[1]	                 \\
     0.547 & $     2.094   \pm  0.023  $  &   {\it Hipparcos} V &[1]                    \\
      1.235 & $     1.342  \pm   0.030  $  &   {\it 2MASS} J &[2]	              \\
      1.65 & $    0.913  \pm   0.044  $  &   {\it 2MASS} H &[2]	          \\
      2.16 & $    0.611  \pm   0.010  $  &   {\it 2MASS} K$_\text{s}$ &[2]	                 \\
      3.4 & $    0.278  \pm  0.010  $  &   {\it WISE} 1 &[3]	             \\
      4.6 & $    0.150  \pm  0.005  $  &   {\it WISE} 2& [3]	               \\
      8.00 & $  0.064 \pm   0.002   $ &  {\it Spitzer} IRS		&     [4]               \\
      8.65 & $  0.059  \pm  0.003   $ &  {\it Spitzer} IRS		&     [4]               \\
      9.35 & $  0.053  \pm  0.003   $ &   {\it Spitzer} IRS		& [4]                   \\
      10.11 & $   0.049  \pm  0.002  $ &    {\it Spitzer} IRS		&      [4]              \\
      10.93 & $   0.049  \pm  0.002  $  &   {\it Spitzer} IRS		&           [4]         \\
      11.2 & $   0.050  \pm  0.002  $  &  Gemini-N/Michelle& [5]                   \\
      11.56 & $   0.053  \pm  0.005  $  &    {\it WISE} 3 & [3]	                  \\
      11.7 & $   0.053   \pm 0.005  $  &  Gemini-S/T-ReCS &[6]	             \\
     11.81 & $  0.050  \pm  0.002   $   &   {\it Spitzer} IRS		& [4]                 \\
      12.77 & $   0.052  \pm  0.001  $    &  {\it Spitzer} IRS		& [4]                  \\
      13.80 & $   0.054  \pm  0.003  $    &  {\it Spitzer} IRS		&  [4]                 \\
      14.92 & $   0.063  \pm  0.002  $    &  {\it Spitzer} IRS		&       [4]            \\
      16.13 & $   0.072  \pm  0.003  $    &  {\it Spitzer} IRS		&            [4]       \\
      17.44 & $   0.087  \pm  0.004  $    &  {\it Spitzer} IRS		&             [4]      \\
      18.3 & $   0.090   \pm  0.014  $  &  Gemini-S/T-ReCS &[6]	               \\
      18.85 & $    0.110  \pm  0.004  $    &  {\it Spitzer} IRS		&    [4]               \\
      20.38 & $    0.142 \pm   0.009  $   &   {\it Spitzer} IRS		&    [4]              \\
      22 & $    0.193  \pm  0.020  $  &   {\it WISE} 4& [3]	                   \\
     22.03 & $   0.189   \pm 0.009   $  &    {\it Spitzer} IRS		&     [4]            \\
      23.68 & $    0.210  \pm   0.010  $  &  {\it Spitzer} MIPS &[7]                  \\
      23.81 & $    0.239  \pm   0.011  $   &   {\it Spitzer} IRS		&      [4]            \\
      25 & $    0.256  \pm   0.041  $  &   {\it IRAS} 25& [8]	          \\
      25.74 & $    0.296  \pm  0.010  $   &    {\it Spitzer} IRS		&  [4]               \\
      27.83 & $    0.375  \pm   0.011  $    &  {\it Spitzer} IRS		&     [4]              \\
      30.08 & $    0.444   \pm  0.011  $    &  {\it Spitzer} IRS		&        [4]           \\
      32.51 & $    0.563  \pm   0.014  $    &  {\it Spitzer} IRS		&           [4]        \\
      35.15 & $   0.668   \pm  0.018   $  &    {\it Spitzer} IRS		&              [4]   \\
      60 & $     1.140   \pm  0.070  $  &  {\it IRAS} 60 &[8]	                    \\
1200		& $0.00314\pm0.00095$	&	IRAM 30m/MAMBO2		& [9] \\
1300		& $0.00310\pm0.00087$	&	SMA					& [9] \\
\enddata
\tablecomments{Color corrected flux with $1\sigma$ error bars used in the SED modeling.  References - [1]: \cite{Hog00}, [2]: \cite{Cutri03}, [3]: \cite{Wright10}, [4]: Spitzer Heritage Archive, [5]: \cite{Fitzgerald07}, [6]: \cite{Moerchen07}, [7]: \cite{Maness08}, [8]: \cite{Moor06}, [9]: \cite{Meeus12}}
\end{deluxetable}

\subsection{Stellar Properties}
We started our analysis by fitting for the stellar parameters of HD32297.  We used the {\it Hipparcos} B \& V, {\it 2MASS} J, H, \& K$_\text{s}$, and {\it WISE} bands 1 \& 2 to constrain the stellar models.  We fit the stellar data with ATLAS9 stellar photosphere models \citep{Castelli04}.  The spectral type of HD32297 is usually quoted as either an A0 \citep{Torres06} or an A5 \citep[][and references therein]{Fitzgerald07}.  Our best fitting model with an A0V spectral type (T = 9750 K) is too hot and does not match the data well.  

We therefore tried both adding interstellar extinction with a \cite{Fitzpatrick99} extinction law and varying the stellar photosphere temperature.  Unfortunately, these two parameters are degenerate.  Fixing the temperature to T = 9750 K, the best fitting model has an extinction of A$_\text{V} = 0.63\pm0.02$ mag, a reduced chi-squared value of $\chi^2_\nu = 0.59$, and a bolometric luminosity of L = 11.9 L$_\odot$. The best fitting model with extinction and temperature as free parameters has T = 8000 K, A$_\text{V} = 0.161\pm0.026$ mag, and $\chi^2_\nu = 0.55$.

To break the degeneracy, we added UV continuum values at 0.26 and 0.31$\,\mu$m from unpublished STIS spectra (Redfield et al.\ 2012, in preparation).  The best model is one with no extinction, a temperature of T=7750 K, and a luminosity of L = 5.6 L$_\odot$ (Fig. \ref{fig:SED}).  This is more consistent with an A7 spectral type than an A0.  This hint in the UV spectra that the spectral type is later than was thought will be thoroughly analyzed in a future paper (Redfield et al.\ 2012, in preparation).  We note that this lower temperature is the same used by \cite{Debes09} and close to the temperature used by \cite{Fitzgerald07}, both of whom note that the star appears to be under-luminous for a main sequence star of this temperature.  Some reasons proposed are errors in the Hipparcos distance \citep{Debes09} or a sub-solar metallicity \citep{Meeus12}.  

The assumed stellar luminosity does not have a significant effect on the SED in the far-IR where the stellar contribution is negligible.  However, the assumed temperature of the star is important for calculating the equilibrium temperature of the dust.  An error in the stellar temperature will translate into an error in the dust temperature, which will affect the calculated grain properties, such as minimum grain size.  

\subsection{Surface Density Profile}

The HD32297 disk has been well resolved in scattered light with HST NICMOS at 1.1$\,\mu$m \citep{Schneider05}, 1.6$\,\mu$m, 2.05$\,\mu$m \citep{Debes09}, and K$_\text{s}$ band ($\lambda = 2.16\,\mu$m) from the ground with Keck/NIRC2 \citep{Currie12}, as well as H ($\lambda = 1.65\,\mu$m) and K$_\text{s}$  bands with VLT/NACO \citep{Boccaletti12}.  These observations block out the central star with a coronagraph, and consequently obscure the inner portions of the disk as well.

The resolved images place strong geometrical constraints on the disk outside of $\sim65$ AU.   \cite{Currie12} and \cite{Boccaletti12} have both recently published models of the disk based on their near-IR ground based imaging using ADI.  \cite{Boccaletti12} warn that images processed using ADI and/or the Locally Optimized Combination of Images (LOCI) are subject to self-subtraction and other artifacts \citep[also see][]{Milli12}.  Therefore, a direct inversion of the surface brightness profile in the manner of \cite{Augereau06} is not practical here.  For this reason, we rely on the disk modeling that takes into account the ADI and LOCI processing.  

The models of \cite{Currie12} and \cite{Boccaletti12} agree quite well.  Both see a break in the surface brightness profiles around $1^{\prime\prime}$ ($\sim 110$ AU), which was also seen by \cite{Debes09} in the NICMOS images.  Both measure an inclination of $88\degr$, i.e.\ $2\degr$ from edge-on.  This is notably different from the inclination measured by \cite{Schneider05} of $79.5\degr$.

Where the models disagree is in the anisotropic scattering factor, $g$.  \cite{Currie12} use a two-component Henyey-Greenstein phase function with a highly forward scattering component ($g_1 = 0.96$) and a backscattering component ($g_2 = -0.1$), while  \cite{Boccaletti12} use only a one-component phase function with a best fit value of $g=0.5$.  They discuss how higher values of $g$ would make the disk too bright in the inner regions.  \cite{Currie12} also note this, but they dismiss it due to the large uncertainties in surface brightness close to the star.  

We chose to use the models of \cite{Boccaletti12} because they used the GR{\footnotesize A}T{\footnotesize ER} code \citep{Augereau99, Lebreton12} to model the disk images, which we also used to model the SED (see Section 4.5).  The best-fit model of \cite{Boccaletti12} to the K$_\text{s}$ band image has a mid-plane density of the form \begin{equation} n(r) = n_0 \sqrt{2} \left( \left( \frac{r}{110 \text{AU}} \right)^{10} + \left( \frac{r}{110 \text{AU}} \right)^{-4} \right)^{-1/2}. \end{equation}  We assumed the disk is geometrically thin because the SED modeling cannot distinguish between a vertical offset from the midplane and a radial change in distance. We chose to use the \cite{Boccaletti12} model of the K$_{\text{s}}$ band image rather than the H band image because the K$_\text{s}$ band data are of better quality.

\subsection{Dust Disk Modeling Strategy}

We started our modeling by fitting the infrared excess with a single temperature modified blackbody, i.e.\ a blackbody model with an opacity at longer wavelengths of the form $\nu^\beta$.  The exact form of the modified blackbody we used is $F_\nu \propto \tau_\nu B_\nu$ for $\lambda \geq \lambda_0$, and $F_\nu \propto B_\nu$ for $\lambda < \lambda_0$, where $B_\nu$ is a simple blackbody and $\tau$ is the optical depth, which takes the form $\tau \propto (\nu/\nu_0)^\beta \propto \left(\lambda/\lambda_0\right)^{-\beta}$. 
Here we have fixed $\beta=1$ and $\lambda_0 = 100\,\mu$m, based on typical values for debris disks \citep{Dent00,Williams06}.  The best fit was determined through $\chi^2$ minimization using \texttt{MPFIT} \citep{Markwardt09}.  
The best fitting model has a temperature of T = 80 K.  However, this model significantly 
underestimates the flux at mid-IR wavelengths. 

To account for the missing IR flux in our model, we added a second modified blackbody.  The best fitting model has two distinct components with temperatures of 
 T = 83 K and T= 240 K (see Fig. \ref{fig:SED}).  The addition of the second blackbody much improves the fit over the single temperature model, suggesting there is a second inner disk.  The inner component is too hot, and therefore too close to the star, to be part of the outer ring imaged in scattered light.
  
Spatial information is important for breaking degeneracies in SED modeling.  Scattered light images of HD32297 restrict the models of the outer disk geometry.  Unfortunately, no constraining images exist for the inner disk, which is too close to the star to be imaged.  Therefore, we divided the dust disk modeling into two steps: first fitting the outer disk with a more complex model, then fitting the poorly constrained inner disk with a simpler one.

\subsection{Outer disk modeling with GR{\footnotesize A}T{\footnotesize ER}}

For the outer disk, we limited the data to those with wavelengths larger than $25\,\mu$m. 
This cutoff was chosen as the point where the inner disk contributes less than 50\% to the total two component blackbody model.  
With the {\it Herschel} observations, we have enough data points with wavelengths greater than $25\,\mu$m to constrain the outer disk properties. 

We use the GR{\footnotesize A}T{\footnotesize ER} code \citep{Augereau99, Lebreton12} to fit the SED of HD32297.  GR{\footnotesize A}T{\footnotesize ER} is dust disk modeling code specifically designed for optically thin debris disks, which can compute large grids of models with different grain sizes and composition and use a density profile to describe the disk geometry.  The code computes both the scattered light emission and the thermal emission from the grains in equilibrium with the radiation field of the central star. 
We constrained the spatial distribution of the dust grains by using the models of the resolved scattered light images from \cite{Boccaletti12}.  We confined the grains in our SED model to be located in a ring defined by Equation 4.  The overall abundance was left as a free parameter that scales the radial profile to match the SED.  

With the disk geometry fixed, we focused our modeling on the grain sizes and composition, which we assumed to be the same throughout the disk.  
GR{\footnotesize A}T{\footnotesize ER} calculates Mie scattering and absorption coefficients for a large range of grain compositions.  Specifically, we explored combinations of materials consisting of astrosilicate \citep{Draine03}, amorphous carbon \citep{Zubko96}, amorphous water ice \citep{Li98} and porosity. The volume ratios of the materials explored are listed in Table \ref{tab:parameters}.  If any material reaches its sublimation temperature, it is replaced by vacuum.  For most of the outer disk, the temperatures are too cold for this to happen.  

For the grain sizes, we explored a range of minimum grain sizes from a$_\text{min} = 0.01\,\mu$m to a$_\text{min} = 100\,\mu$m and grain size distributions with a power-law of the form $n(a)da\propto a^\kappa da$, with $\kappa$ ranging from $-5$ to $-2.5$.  The maximum grain size is kept fixed at a$_{\text{max}} = 7.8$ mm, which is large enough compared to the longest wavelength of the data ($\lambda = 1.3\,$mm).  However, we only considered grains with sizes smaller than 1 mm in calculating the dust mass (given in Table \ref{tab:results}), to be consistent with other results from GASPS and other {\it Herschel} Key Programmes.  

We modeled the disk using six different compositions with varying levels of complexity.  The simplest model used only  astrosilicate grains.  The next three models used silicates mixed with only one other grain type: carbonaceous material, water ice, or increased porosity.  For the three and four material combinations, we kept the silicate to carbon volume ratio fixed at 1:2.  This is the ratio expected from cosmic abundances and is similar to the ratio observed in comet Halley dust \citep{Greenberg98}.  

The grains are assumed to be porous aggregates of silicate, carbon and water ice.  The scattering and absorption coefficients for the aggregates are calculated using the Bruggeman mixing rule.  Silicate and carbon are mixed first, then the Si+C mixture is mixed with water ice, and finally it is all mixed with vacuum to simulate porous grains.  For more details see \cite{Lebreton12} and \cite{Augereau99}.  Grain densities for the mixtures appear in Table 5.  

\begin{table}[ht]
\centering
\caption{Parameters explored in GR{\tiny A}T{\tiny ER} models \label{tab:parameters}}
\begin{tabular}{l c c c}
\tableline \tableline 
Parameter		& range	& \# of values	& distribution \\
\tableline 
$\kappa$		& -5 to -2.5 &	20		& linear 	\\
a$_\text{min}$	& 0.01 to $100\,\mu$m & 77	& logarithmic \\
Carbon Volume	 & 0 to 100\% 	& 21		& linear \\
Ice Volume	& 0 to 90\%	& 10		& linear \\
Porosity		& 0 to 95\%	& 20		& linear \\
\tableline  
\end{tabular}
\end{table}

\subsection{Inner disk modeling}

The mid-IR flux can only be explained with a warmer component than is seen in the scattered light images. We first tried to model the disk with the geometry constrained only by the coronagraphic scattered light images of the disk. We found that the thermal emission from the grains seen in the scattered light images was not enough to reproduce the mid-IR flux in the SED.   The data in this region come from several sources, and all agree with each other ({\it Spitzer} IRS, {\it WISE} \citep{Wright10}, and Gemini N \citep{Fitzgerald07} and S \citep{Moerchen07}).  Therefore, we added a warmer component to fit the mid-IR data.  

We chose to model the warm component as an inner disk inside the radius masked by coronagraphs in the images.  The inner disk is less constrained than the outer disk since the inner disk lacks geometrical information and the L$_\text{IR}$/L$_\ast$ of the inner disk is an order of magnitude lower than the outer disk ( 6.9$\times$10$^{-4}$ vs. 5.6$\times$10$^{-3}$; see Fig. \ref{fig:bestfit}).  This leads to a degeneracy between two of the dust disk modeling parameters: the minimum radius and the minimum grain size.  The degeneracy exists because decreasing both the parameters has the same effect of increasing the grain temperature.  Changing the minimum grain size also affects the amount of flux emitted from a given grain, but this can be mimicked by varying the total number of grains.  With a weakly-emitting disk, the separate effects can be difficult to disentangle. 

For each outer disk model with a different grain composition, we subtracted it from the SED and fit our inner disk model to the residuals.  Since we lack constraining spatial information for the inner disk, we needed to use a simpler model.  We used the model described in \cite{Donaldson12}, which calculates only the thermal emission from astrosilicate grains in radiative equilibrium with the central star.   We fixed the outer radius to 5 AU since all values above this had no significant effect on the SED.  We assumed the disk has a surface density profile of the form $\Sigma (r) \propto r^{-1.5}$, consistent with collision-dominated disks \citep{Krivov06, Strubbe06}.  We also fixed the grain size distribution throughout the disk to a \cite{Dohnanyi69} power-law ($n(a) da \propto a^{-3.5} da$) with a maximum grain size of 1 mm.  The only free parameters were the inner radius, the minimum grain size, and the dust mass.  

We used the inner disk model to fit the residuals after subtraction of the outer disk model, then combined this model with those of the outer disk and the star.  We calculated reduced $\chi^2$ values ($\chi^2_\nu$) using all 43 data points listed in Tables \ref{tab:photometry}, \ref{tab:spec}, and \ref{tab:archive} greater than $8\,\mu$m and the number of degrees of freedom ($\nu$) listed in Table \ref{tab:results}.  We calculated the errors on the parameters from the $1\sigma$ confidence intervals in the $\chi^2$ distribution.   

A similar modeling approach was recently used by \cite{Ertel11} for HD107146.  They fit the disk SED using the scattered light images to constrain the geometry, and they also found an overabundance of {\it Spitzer} IRS mid-IR flux in their best fit models.  They tried modeling the excess in two ways: first by adding a small grain population within the imaged disk, and second by adding an inner disk.  The small grain model was unable to reproduce the mid-IR flux in HD107146; the inner disk model was needed to match the {\it Spitzer} IRS spectrum.

\subsection{Results}

\begin{figure*}
\epsscale{1.2}
\plotone{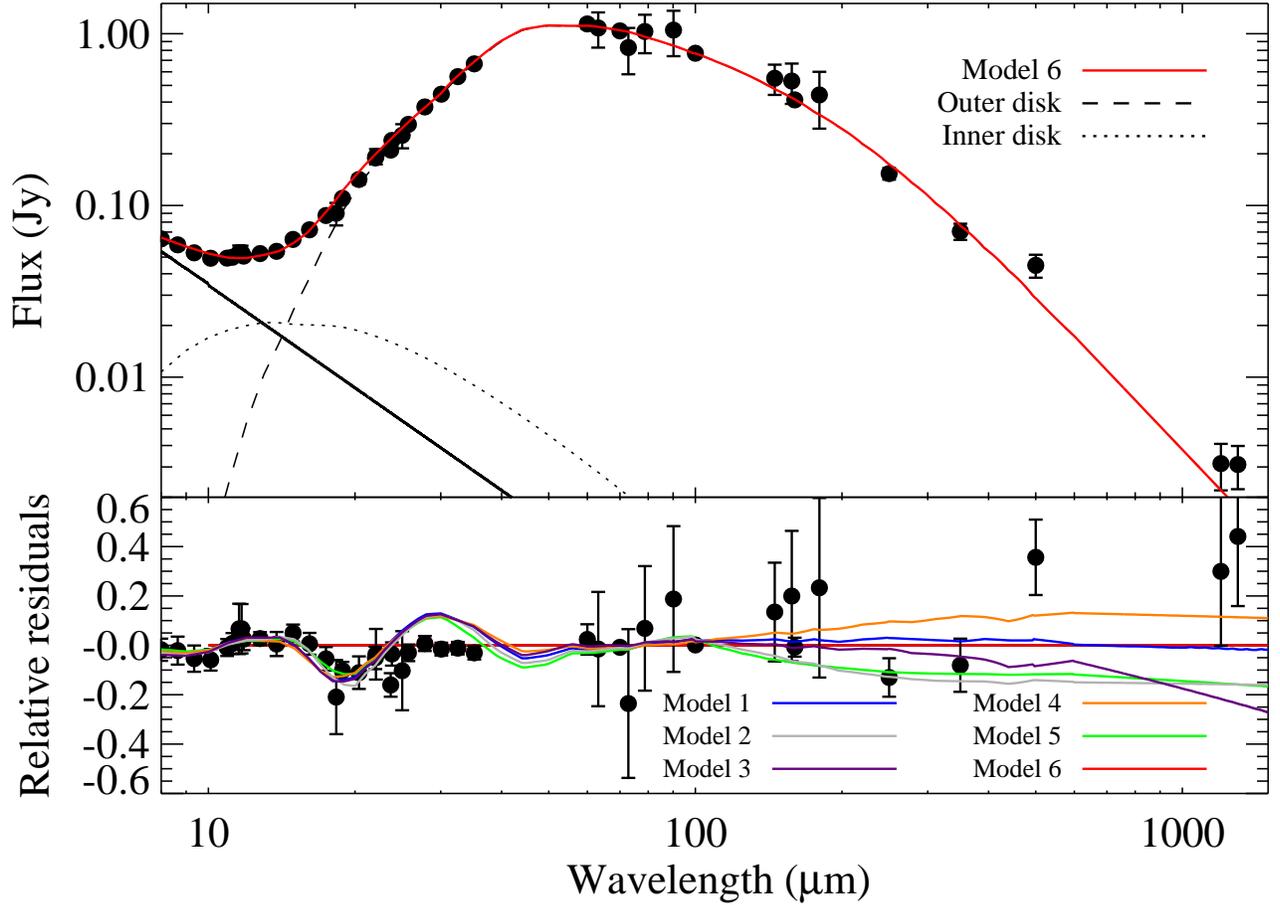}
\caption{Top: Spectral energy distribution for the best-fitting total model given in Table \ref{tab:results} (Model 6).  The black solid line is the fit to the stellar photosphere, the dotted and dashed lines show the inner and outer disk models respectively, and the red solid line represents the combined model of photosphere + inner disk + outer disk. The data plotted here are listed in Tables \ref{tab:photometry}, \ref{tab:spec}, and \ref{tab:archive}.  Bottom: Bottom: The relative residuals of the best-fitting model shown above. The relative residuals are of the form (data - model\#6)/data.  Also plotted for comparison are the relative residuals of the other five models given in Table \ref{tab:results}, e.g.\ (model\#1 - model\#6)/model\#1. The red line marks the zero point, or the residual for model 6.   \label{fig:bestfit}}
\end{figure*}

We found the best fitting outer disk model was the 4 material composition of silicates, carbon, and water ice in a 1:2:3 ratio with a high porosity of 90\% (final row of Table \ref{tab:results}, Fig. \ref{fig:bestfit}). The outer disk grains have a minimum size of $a_{min} = 2.1\,\mu$m, with a grain size distribution power-law index of $\kappa=-3.3.$  The best fitting total SED model also includes an inner disk from 1.1 AU with an unconstrained outer edge.  The inner disk was fit with astrosilicate grains with a \cite{Dohnanyi69} size distribution ($\kappa=-3.5$).  The minimum grain size in the inner disk for the best fitting model is  $a_{min}=2.2\,\mu$m, similar to the outer disk grains.  Other grain models tested appear in Table \ref{tab:results}.  We determined the uncertainties given in Table \ref{tab:results} from the $1\sigma$ confidence intervals in the $\chi^2$ distribution after fixing the other free parameters to the value that gives the smallest $\chi^2$ value.  

The 4 material composition model has the best reduced $\chi^2$ value of the tested compositions ($\chi^2_\nu = 1.59$, $\nu=34$).  Of the 2 material composition models, astrosilicate + carbon and  astrosilicate + porosity are the best fitting ($\chi^2_\nu = 4.27$ and $\chi^2_\nu = 4.42$, $\nu=35$), but the model is much improved when all 3 materials are added together ($\chi^2_\nu = 3.60$, $\nu=35$).  The addition of ice makes a dramatic improvement in the fit from the 3 material composition to the 4 material composition.  The minimum grain size in the inner disk models decreases with each improvement on the outer disk model, moving closer to the expected blowout size of $1\,\mu$m.  This is likely because the two material models fail to fit all the mid-IR flux coming from the outer ring, and the inner disk models have to compensate.  

The blowout size given above is calculated for the inner disk only, where the assumed grain composition is astrosilicate.  Given a density of $\rho=3.5$ g cm$^{-3}$, the blowout size for spherical grains on circular orbits (in microns) is a$_\text{blow} = 1.15 \text{L}_\ast / (\text{M}_\ast \rho)$, with $\text{L}_\ast$ and $\text{M}_\ast$ in solar units and $\rho$ in g cm$^{-3}$.  For astrosilicate grains around HD32297 ( M$_\ast$ = 1.84 M$_\odot$, L$_\ast$ = 5.3 L$_\odot$), the blowout size is $a_\text{blow} = 1\,\mu$m.  

For the outer disk, the grains are highly porous and likely have a fractal structure.  The calculation of the blowout size depends on the surface area of the grains; this is non-trivial for fractal grains.  The above equation is for spherical grains only; we do not calculate the blowout size for the outer disk because it is not very realistic in this case.  For more about this problem, see the discussion in \cite{Lebreton12}.  

\begin{deluxetable*}{l c c c c c c c c c c c}
\tablecolumns{11}
\tabletypesize{\scriptsize}
\tablecaption{Results of SED modeling\label{tab:results}}
\tablehead{&volume&density&\multicolumn{2}{c}{outer disk} && \multicolumn{2}{c}{inner disk} && \multicolumn{3}{c}{total disk} \\
\cline{4-5} \cline{7-8} \cline{10-12}
Outer Disk Composition\tablenotemark{a} & ratios 	& (g cm$^{-3}$) & a$_{\text{min}}$	& $\kappa$	&& r$_{\text{min}}$	& a$_{\text{min}}$	&& $\chi^2_\nu$ &$\nu$ & Dust Mass\tablenotemark{b}\\}
\startdata
Model 1 (AS)	& --	& 3.50& $1.4\pm0.1\,\mu$m	& $-3.8\pm0.2$	&& $3.3\pm0.3$ AU 	& $200\pm120\,\mu$m	&& 4.58	& 36		& $0.32\pm0.05$ M$_\oplus$	\\
Model 2 (AS+C) &	1:4	& 2.46& $2.1\pm0.4\,\mu$m	& $-4.3\pm0.3$	&& $3.5\pm0.5$ AU & $31.6\pm26\,\mu$m	&&	4.27	& 35 & $0.11\pm0.02$ M$_\oplus$\\
Model 3 (AS+P)  & 11:9 &1.93	& $0.3\pm0.1\,\mu$m	&$-3.5\pm0.2$	&&$3.6\pm0.4$ AU	& $31.6\pm27\,\mu$m	&& 4.42	& 35 & $0.26\pm0.05$ M$_\oplus$ 		\\
Model 4 (AS+I) & 4:1 &3.04	& $1.0\pm0.1\,\mu$m	& $-3.7\pm0.2$	&&$3.2\pm0.3$ AU	&$39.8\pm22\,\mu$m	&& 4.45	& 35&$0.34\pm0.12$ M$_\oplus$	\\
Model 5 (AS+C+P) & 1:2:12&0.53	& $3.4\pm0.1\,\mu$m	& $-3.8\pm0.2$	&& $2.8\pm0.3$ AU		& $25.1\pm24\,\mu$m	&& 3.60	& 35 & $0.08\pm0.01$ M$_\oplus$ \\
Model 6 (AS+C+I+P)& 1:2:3:54 &0.19	& 	$2.1\pm0.3\,\mu$m	& $-3.3\pm0.2$	&& $1.1\pm0.2$ AU	& $2.2\pm0.9\,\mu$m	&& 1.59	& 34 & $0.10\pm0.01$ M$_\oplus$\\
\enddata
\tablenotetext{a}{Model 1: Astrosilicate (AS), Model 2: Astrosilicate + Carbon (C), Model 3:  Astrosilicate + Vacuum (P), Model 4:  Astrosilicate + Water Ice (I), Model 5: Astrosilicate+ Carbon + Vacuum, Model 6:  Astrosilicate + Carbon + Water Ice + Vacuum}
\tablenotetext{b}{The dust mass is calculated for dust grains smaller than 1 mm only.  The total dust mass is dominated by the outer disk.}
\end{deluxetable*}

\section{Discussion}
The composition of dust grains in young debris disks is a key piece in understanding the last stages of terrestrial planet formation.  A handful of debris disks, including $\beta$ Pictoris and HD172555 \citep{Telesco91,Lisse09}, have solid state features in the mid-IR that indicate dust grain composition.  But most debris disks, including HD32297, lack these features, and therefore, modeling of the full SED is needed to constrain the grain composition.

Unfortunately, the presence of an unresolved warm component to the HD32297 system complicates the modeling.  Without resolved imaging of the inner regions, it is impossible to know the distribution of the warm component and how much mid-IR flux is coming from the outer disk versus the inner disk.  The models of the inner disk depend strongly on the model chosen for the outer disk.  Additionally, since there are no geometrical constraints on the inner disk, the results depend upon the distribution we assumed.  

Another concern is the age of the system.  \cite{Kalas_30Myr} states the age as 30 Myrs based on an uncertain association with either the Gould belt or Taurus Aurigae.  The dust mass of HD32297 (see Table \ref{tab:results}) is high for a debris disk.  The mass is comparable to the 8 Myr-old disk HR4796A, which suggests HD32297 may be younger than 30 Myrs.  HD32297 is also one of the oldest debris disks with gas detected, another indication it may be a younger system.  But the system likely not much younger than the given age of 30 Myrs.  There are several indicators that HD32297 is a main sequence star with an optically thin debris disk, including 1) the lack of optical extinction, 2) the low fractional dust luminosity (L$_\text{IR}/L_\ast \sim 10^{-3}$) and 3) a lack of solid state features in the mid-IR indicating no small grains are present.  

\subsection{Cometary Dust?}
The best fitting model to the outer disk includes grains that are highly porous and icy.  High grain porosity is seen in interplanetary dust particles collected with Stardust \citep{Brownlee06} and in the ejecta from comet Temple 1 created by Deep Impact \citep{A'Hearn08}.  \cite{Greenberg90} showed that comet Halley's spectrum could only be fit by highly porous grains with a porosity between 93 and 97.5\%.
\cite{Li98} modeled the $\beta$ Pictoris disk with similar composition dust.  They assumed the $\beta$ Pic dust was cometary in origin and rejected models of compact grains with porosity lower than 90\%.  Polarized light observations of AU Mic also indicate that disk is dominated by highly porous grains with porosity of 91-94\% \citep{Graham07,FitzgeraldAUMic}. 
The highly porous, icy dust around HD32297 is similar to $\beta$ Pic and Solar System comets.  

Dust in the outer ring of HD32297 therefore appears consistent with cometary dust particles.  
The ring is centered around $\sim110$ AU, far from the star where ices should be prevalent.  If planetesimal collisions produce the dust, this indicates that comet-like bodies dominate the planetesimal population in the outer disk.  

A large population of comets in the outer disk could deliver water to terrestrial planets.  A significant fraction of Earth's water likely came from Kuiper Belt comets \citep{Morbidelli00}.  At an age of 30 Myrs, HD32297 could still be forming terrestrial planets \citep{Kenyon06}.  Comets scattered into the inner regions of the disk could deliver water to forming terrestrial planets in the habitable zone.  

\subsection{Grain Porosity and ISM interaction} 

The scattered light images of HD32297, as well as those of the edge-on disks HD15115 and HD61005, have asymmetric features that may be due to interaction with ISM as the systems move with respect to their surroundings.  The short wavelength of images of HD32297 \citep{Schneider05, Debes09} and HD61005 \citep{Hines07, Maness09, Buenzli10} show a ``swept-out'' feature, while HD15115 has a strong east-west asymmetry \citep{Kalas07, Debes08, Rodigas12}.   The ISM interaction model of \cite{Debes09} reproduces the features of all three disks.   

The ISM affects the disk grains through gas drag and/or grain-grain collision as the disk moves through a dense clump in the ISM.  Unbound or weakly bound grains are swept back as they interact with the ISM.  The most affected grains are typically thought to be the small grains that are nearly unbound due to the effect of radiation pressure.  But larger grains with a higher porosity are also strongly affected by radiation pressure, and therefore may also be susceptible to being blown back through ISM ram pressure.  The high porosity of the outer disk grains may be one factor that helps explain why this disk has such a strong ISM interaction feature.  Stellar motion and environment must also play a role, since other disks modeled with a high grain porosity, such as $\beta$ Pic, do not have the same feature.  

\subsection{Gas in HD32297}
HD32297 is one of only a handful of debris disks that have gas detections.  \cite{Redfield07} found \ion{Na}{1} in absorption, aided by the disk's nearly edge-on orientation.  The detection of [\ion{C}{2}] emission from HD32297 is the fourth detection of atomic gas from a debris disk with {\it Herschel}, though it is weaker than the lines seen from $\beta$ Pictoris, HD172555, and 49 Ceti \citep{Brandeker12,Riviere-Marichalar12, Roberge12}.  
It is also unusual that [\ion{C}{2}] was detected while [\ion{O}{1}] was not.  The only other debris disk with gas where this is true is 49 Ceti \citep{Roberge12}.  

Given relatively advanced age of HD32297 \citep[$\sim30$ Myr;][]{Kalas_30Myr} and the typical protoplanetary disk lifetime \citep[$<10$ Myrs;][]{Williams11}, the HD32297 gas is unlikely to be primordial.   The lack of sub-mm CO emission suggests a gas-to-dust ratio lower than is seen in younger protoplanetary disks \citep[e.g.][]{Zuckerman95}.  Furthermore, the disk dust has a relatively low abundance, and shares other characteristics with debris dust (no detectable line-of-sight extinction, a lack of mid-IR solid state features from small grains).  With little dust and little gas, the disk should be optically thin to stellar and interstellar dissociating UV photons and molecular gas lifetimes should be short.  However, at this time, it is difficult to prove that the observed \ion{C}{2} is not simply the tenuous end product of dissociated primordial gas, although the lack of \ion{O}{1} emission would be puzzling in this scenario.  An alternative scenario would be gas production by secondary mechanisms such as planetesimal collisions or outgassing from comet-like bodies.

To calculate the total amount of carbon gas in the disk, we assume the disk is similar to the well studied $\beta$ Pictoris debris disk.  Several gas species have been observed in $\beta$ Pic, and their abundance ratios are summarized in \cite{Roberge06}.  The ratio of neutral and singly ionized carbon in $\beta$ Pic is \ion{C}{1}/\ion{C}{2} $\sim$ 1.  Assuming the same ionization fraction, the total column density of carbon in HD32297 is N$_\text{C} \gtrsim 5\times10^{11} \text{ cm}^{-2}$.  The ionization fraction of sodium in $\beta$ Pic is $\geq 0.999$ \citep{Roberge06}.  This implies a total sodium column density of N$_{\text{Na}} > 2.5\times10^{14}$ cm$^{-2}$, 500 times larger than the lower limit on carbon.  

By assuming solar abundances, we naively expect there to be about two orders of magnitude more carbon than sodium.  This would only be the case if the excitation temperature is less than 10 K.  We consider three possible explanations.  The first possibility is that the excitation temperature really is less than 10 K;  this would be several times lower than the excitation temperature measured in $\beta$ Pic \citep{Roberge06}.  Second, the disk does not have solar abundances as assumed, but a different ratio, meaning less carbon or more sodium.  Yet, we expect the opposite to the true. Carbon does not feel as strong a radiation pressure as sodium because unlike sodium, carbon does not have strong absorption lines in the optical, but in the far-UV where the star is much fainter \citep{Roberge06}.  Hence, we expect there to be more carbon than sodium relative to solar abundances, making the problem worse.  The last possibility is that ISM sodium absorption lines along the line of sight contaminated the HD32297 sodium measurements, boosting the signal.  We deem this to be the most likely scenario.

We can estimate a lower limit on the total gas mass in HD32297 by making a few assumptions.  We start by modifying Equation 1 to get a \ion{C}{2} mass lower limit of M$_\text{CII} > 1.7 \times 10^{-4} \text{ M}_\oplus$.  By making the same assumptions as above (\ion{C}{1}/\ion{C}{2}$\sim1$, solar abundances), the total gas mass is $\sim700$ times the \ion{C}{2} mass, giving a lower limit of M $> 0.1 $ M$_\oplus$.

Gas in disks can affect the distribution of dust. Gas orbits the star at sub-keplerian speeds due to either to a gas pressure gradient or radiation pressure.  The dust grains, if large enough not to experience strong radiation pressure, orbit at keplerian speeds, and therefore feel a headwind that causes them to spiral inwards.  This mechanism could be a way of transporting dust grains from the outer disk to the inner disk.  \cite{Krivov09} investigated how 0.3-30 M$_\oplus$ of gas would affect HD32297's radial surface brightness profiles, and found the gas has little effect on the disk outside 110 AU.  The distribution of the inner disk, however, could be affected by small amounts of gas.  Since we have no limiting spatial information on the gas in HD32297, we cannot determine if the gas significantly affects the distribution of the dust in the inner disk.  This scenario will be further discussed in Section \ref{sub:inner}.

\subsection{Inner Disk\label{sub:inner}}

Our best model for the inner disk of HD32297 starts at $\sim1$ AU.  The outer edge of the inner disk is unconstrained, but we find a lower limit on the outer edge of $\sim5$ AU.  We assumed the inner disk grains are astrosilicate grains, and our best fitting model has a minimum grain size of $2\,\mu$m.  The fit to the inner disk depends strongly on the fit to the outer disk.  Although the results of the grain size in the inner disk varied by about 2 orders of magnitude (2.2$\,\mu$m - 200$\,\mu$m) the inner radius only varied by a factor of a few (1.1 - 3.6 AU).  This range places the inner disk near the habitable zone of the star.  A simple $\sqrt{L_\ast}$ scaling of the Solar System's habitable zone ($\sim0.7$ - 1.5 AU) places the HD32297 habitable zone at $\sim 1.7 - 3.5$ AU. 

The presence of dust in the habitable zone does not rule out terrestrial planets in this region.  Low mass planets may not have had time to clear this region of planetesimals.  In fact, some of the dust may be trapped in resonance with a planet \citep{Wyatt03,Stark08}. 

Since the inner disk has not been resolved, we do not know how the dust is distributed.  The dust distribution in the inner region depends on the location of the parent material and dust transportation.  We consider two scenarios for the origin of the inner dust disk: one where the inner dust disk is fed by dust transported from the outer disk through gas drag, and the second where there is another planetesimal belt closer to the star.  

In the first scenario, the presence of gas might affect the dust distribution, such that it is no longer collision-dominated.  The entire inner disk may be composed of material that has leaked inwards from the outer disk due to gas drag.  If this were the case, we would expect a smooth surface density distribution from the outer disk to the inner disk.  At first glance, this might seem inconsistent with an SED that is well fit by a two temperature blackbody model.  This model is most easily interpreted as two rings with a gap in between.  However, a gap is not needed to produce such an SED.  The region closest to the star will be much warmer and thus brighter than the dust in the intermediate region.  The signature of dust in the intermediate region would be hard to detect in the SED alone.  \citet{Reidemeister11} have shown that in the Eps Eri disk, even a bimodal SED curve can be reproduced with models that assume transport of dust from the outer disk (in that case, caused by stellar winds rather than gas), and thus a continuous distribution of dust from the outer to the inner region.  Since the mass of the inner disk that could account for the observed warm emission is M$_\text{inner} \ga 6 \times 10^{-9} $ M$_\oplus$, and assuming the radius of the inner disk of $\sim$ 1 AU, such a continuous distribution within $\sim$ 100 AU would imply roughly $6 \times 10^{-5}$ M$_\oplus$ worth of dust. For gas drag to work, the gas mass should exceed the dust mass.  Therefore, 0.1 M$_\oplus$ of gas, which is a lower limit that we placed from the [\ion{C}{2}] observations in Sec. 5.4, would be sufficient for the transport. However, without images of the inner $\sim$ 50 AU of the disk, we cannot confirm gas drag as the origin of the inner dust.

In our second scenario for the origin of the inner dust, there could be another belt of planetesimals closer to the star, similar to the asteroid belt.  Collisions between planestimals, in this case, would produce the dust seen in the SED, just like in the outer belt. We made calculations with the model of \citet{Loehne08} and a velocity-dependent critical fragmentation energy from \citet{Steward09}. They suggest that an asteroid belt of a sub-lunar mass at 1.1 AU in a 30 Myr-old system can easily sustain  $10^{-7} M_\oplus$ to $10^{-8} M_\oplus$ worth of dust through a steady-state collisional cascade. This is more than the mass of the inner disk ($M \ga 6 \times 10^{-9} M_\oplus$). Without spatial information on the inner disk, we cannot tell if the presence of warm dust is due to an asteroid belt or from material leaked inwards from the outer disk. In any case, the mass of the inner disk is significantly less than the outer disk; only its proximity to the central star makes it outshine the outer disk in the mid-IR.

\section{Summary}
We present new {\it Herschel} PACS and SPIRE photometry and spectroscopy of the edge-on debris disk around HD32297.  Our main conclusions are the following:
\begin{enumerate}
\item We detected the disk at 13 wavelengths from 63 to 500$\,\mu$m, filling in a gap in the SED in this region. The new data probe the peak of the thermal emission.
\item We detected a 3.7$\sigma$  [\ion{C}{2}] line at 158$\,\mu$m, making HD32297 only the fourth debris disk with atomic gas detected with {\it Herschel}.  We estimate a lower limit on column density of N$_{\text{CII}}> 2.5\times10^{11}$ cm$^{-2}$.  
\item The stellar fit to the optical, near-IR and UV data suggest the star has a later spectral type than typically quoted, likely an A7.  
\item Our SED models require a warm component to fit the large mid-IR excess.  This material is too warm to be part of the ring imaged in scattered light.  The geometry of the warm component is unconstrained; we were able to fit it with a low density disk at radii greater than 1 AU.  
\item The best fitting model to the outer disk includes grains consisting of silicates, carbonaceous material, water ice, and a highly porous structure.  These grains are similar to cometary grains found in the Solar System.  
\end{enumerate}

\acknowledgements
This work is based on observations made with Herschel, a European Space Agency Cornerstone Mission with significant participation by NASA. Support for this work was provided by NASA through an award issued by JPL/Caltech. JL and JCA thank the PNP-CNES for financial support.  We also wish to thank the anonymous referee for a very thorough review that helped improve this paper.

\bibliographystyle{apj}
\bibliography{./tuchor,./draft_v2,./prop,./HD32297bib}

\end{document}